\documentclass[12pt,a4paper]{article}

\usepackage{amsmath}
\usepackage{bm}
\usepackage{amssymb,amsfonts}
\usepackage[vcentermath,enableskew]{youngtab}

\renewcommand{\i}{{\imath}}
\newcommand{\bi}{{\bar{\imath}}}
\renewcommand{\j}{{\jmath}}
\newcommand{\bj}{{\bar{\jmath}}}

\newcommand{\ione}{{\bar{\imath}_1}}
\newcommand{\itwo}{{\bar{\imath}_2}}
\newcommand{\ik}{{\bar{\imath}_k}}
\newcommand{\scdots}{\scriptstyle{\cdots}}
\newcommand{\weevdots}{\scriptstyle{\vbox{\hbox{.}\vskip 1pt\hbox{.}\vskip 1pt\hbox{.}}}}

\newcommand{\ID}{\mathbf{1}}

\newcommand{\CP}{\mathbb {C}\mathbb{P}}

\newcommand{\tA}{\widetilde A}
\newcommand{\K}{\hat{K}}

\newcommand{\calL}{{\mathcal{L}}}
\newcommand{\calA}{{\mathcal{A}}}
\newcommand{\calS}{{\mathcal{S}}}
\newcommand{\calP}{{\mathcal{P}}}
\newcommand{\calN}{{\mathcal{N}}}

\newcommand{\F}{{\mathcal F}}
\newcommand{\D}{{\cal D}}
\newcommand{\Dirac}{\kern 2pt \raise 1pt\hbox{$\slash$} \kern -9pt D}
\newcommand{\bL}{{\bf L}}
\newcommand{\ttL}{{\tt L}}
\newcommand{\ttR}{{\tt R}}

\newcommand{\vectorindex}[1]{\ensuremath{\bm{#1}}}

\newcommand{\ket}[1]{\left| \! \right. #1\left. \! \right\rangle}
\newcommand{\bra}[1]{\left\langle \! \right. #1\left.\! \right| }
\newcommand{\braket}[2]{\left\langle \! \right. #1 \left| \! \right. #2\left. \! \right\rangle}
\newcommand{\ketbra}[1]{\ket{#1}\bra{#1}}

\newcommand{\eq}{\begin{equation}}
\newcommand{\qe}{\end{equation}}

\newcommand{\al}{\alpha}
\newcommand{\bt}{\beta}
\newcommand{\g}{\gamma}
\newcommand{\dl}{\delta}

\newcommand{\avec}{\ensuremath{\bm{\alpha}}}
\newcommand{\bvec}{\ensuremath{\bm{\beta}}}
\newcommand{\svec}{\ensuremath{\bm{\sigma}}}
\newcommand{\tauvec}{\ensuremath{\bm{\tau}}}

\newcommand{\gvec}{\ensuremath{\bm{\gamma}}}

\newcommand{\vv}[1]{\ensuremath{\bm{#1}}}

\newcommand{\non}{\nonumber}

\newcommand{\ad}{a^\dagger}
\newcommand{\Ad}{A^\dagger}

\def\tyng(#1){\hbox{\tiny$\yng(#1)$}}

\newcommand{\comment}[1]{}

\begin{document}

\title{\vspace*{-0.1cm}\hfill \raise 20pt \hbox{\small DIAS-STP-07-19}\\
A universal Dirac operator and noncommutative spin bundles over fuzzy complex projective spaces}
\author{Brian P. Dolan,$^{1,2}$\footnote{email: bdolan@thphys.nuim.ie}~
Idrish Huet,$^{1,3}$\footnote{email: mazapan@stp.dias.ie}\\
Se\'an Murray $^{1,2}$\footnote{email: smury@stp.dias.ie }\
\; {and}\;
Denjoe O'Connor$^1$\footnote{email: denjoe@stp.dias.ie}
\\
\\
$^{\rm 1}$ \it \normalsize School of Theoretical Physics,
Dublin Institute for Advanced Studies,\\
\it \normalsize 10 Burlington Road, Dublin 4, Ireland.\\
\\
{\it \normalsize $^{\rm 2}$ Department of Mathematical Physics}, \\
{\it \normalsize NUI Maynooth, Maynooth, Co. Kildare, Ireland.}\\
\\
{\it \normalsize $^{\rm 3}$ Depto de F$\acute{\i}$sica},
{\it \normalsize Centro de Investigaci\'on y de Estudios Avanzados del IPN}, \\
{\it \normalsize Apdo. Postal 14-740, 07000 M\'exico D.F., M\'exico.}
\\
\date{November 8, 2007}
}
\maketitle
\begin{abstract}
We present a universal Dirac operator for
noncommutative spin and spin$^\mathrm{c}$ bundles over fuzzy
complex projective spaces.  We give an explicit construction of
these bundles, which are described in terms of finite
dimensional matrices, calculate the spectrum and explicitly exhibit the Dirac eigenspinors.  To our knowledge the spin$^\mathrm{c}$ spectrum for $\CP^n$
with $n\ge 3$ is new.
\end{abstract}
\vfill\eject

\section{Introduction}

There has been a growing interest in recent years in noncommutative geometry and its applications to physics, especially quantum field theory \cite{Madore:2000aq} and string theory \cite{Seiberg:1999vs, Myers:1999ps}.
This paper focuses on a particular kind of noncommutative geometry
and associated spaces, known as fuzzy spaces, which serve as regulators in field theories \cite{Grosse:1995ar}. The advantage of this approach over the traditional lattice discretisation lies in the fact that fuzzy spaces retain the isometries of the underlying space.
Fuzzy spaces are finite approximations to suitable continuum manifolds: more precisely a fuzzy space can be defined by a sequence of finite dimensional algebras $\mathcal{A}_L$ that approximate the commutative algebra of functions on the manifold $\mathcal{M}$ as $L \to \infty$,  together
with some differential geometric information such as a Laplacian or a Dirac operator.
The archetypical algebra $\mathcal{A}_L$ being the full matrix algebra of size $d_L$, $Mat_{d_L}(\mathbb{C})$.

So far several fuzzy spaces have been constructed since the introduction of the seminal example, the fuzzy 2-sphere \cite{Berezin:1974du,Hoppe:1982,Madore:1991bw} and various field theories have been studied within this approach both analytically \cite{O'Connor:2003aj, O'Connor:2007} and numerically \cite{Martin:2004un, GarciaFlores:2005xc, Panero:2006bx, O'Connor:2006wv}.
In particular, fuzzy complex projective spaces, flag manifolds and toric varieties have been
constructed \cite{Balachandran:2001dd, Murray:2006pi, Saemann:2006gf} and provide a wide generalization of previous examples.
Important matters to be addressed are how to construct fuzzy spinor fields and the related Dirac operator. These were first constructed in \cite{Grosse:1994ed} for the fuzzy 2-sphere; previous work aimed at generalizing this construction to other fuzzy spaces include \cite{Balachandran:2001CP2,Balachandran:2002bj} and quantization of vector bundles was considered in \cite{Hawkins:1997gj, Hawkins:1998nj}.

In the present work we construct a universal Dirac operator on fuzzy ${\CP}^n$ and its associated eigenspinors; whereas only ${\CP}^{2n+1}$ admits a spin structure, ${\CP}^n$ admits a countably infinite number of spin$^{c}$ structures \cite{Lawson:1990}. We also reproduce the correct spectrum of the commutative Dirac operator up to a truncation.

In noncommutative geometry sections of the spin bundles are replaced with finitely generated projective modules over a noncommutative algebra. This definition of noncommutative sections is motivated by the Serre-Swan theorem \cite{Swan:1962}
which states that the space of smooth sections of a vector bundle over a compact space M is a projective module of finite type over the algebra $\mathrm{C}^\infty(M)$ of smooth functions over M, and any finite projective $\mathrm{C}^\infty(M)$-module can be realized as the module of sections of some bundle over M.  In \cite{Dolan:2006tx}
we provided a construction of fuzzy vector bundles in this spirit
and in the following we extend this to spinor bundles.

Section \ref{review} contains the basic definitions and presents the multi-oscillator construction previously introduced in \cite{Dolan:2006tx}. It also introduces appropriate covariant derivatives used in the definition of the Dirac operator.
Section \ref{spinors} is a brief review of the construction of spin$^{c}$ bundles over ${\CP}^n$.
Section \ref{noncommutative spinors} presents fuzzy spin bundles and
the universal Dirac operator, along with its eigenspinors and a novel presentation of the spectrum and comments about the zero modes. The explicit calculation of the complete spectrum and the
construction of a basis for all fuzzy spinors is left to the appendices.
Finally, in section 5 we briefly discuss charge conjugation.
Our conclusions and outlook are present in section \ref{conclusions}.

\section{Review of Multi-Oscillator Construction}
\label{review}
In \cite{Dolan:2006tx} we introduced a Fock space, $\F^{\mathrm{Total}}$ generated freely by the action of $n(n+1)$ oscillators $(a^\dagger)^\alpha_\i={\ad}^\al_\i=(a^\i_\al)^\dagger$ satisfying
\eq [a^\i_\al, {\ad}^\bt_\j]=\dl^\i_\j \dl_\al^\bt, \quad \i=1,...,n,\quad
\al=1,...,n+1~. \qe
These oscillators, ${\ad}^\al_\i$, carry the anti-fundamental representation of $u(n+1)$ and the fundamental representation of $u(n)$. The generators of $U(n+1)$ and $U(n)$ are given by the Schwinger-Jordan construction as
\begin{align}
\hat{J}^\al{}{}_{\bt}&:={\ad}^\al_\i a^\i_\bt\\
\hat{J}_\i{}^{\j} &:= {\ad}^\al_\i a^\j_\al
\end{align}
respectively. These include a common $U(1)$ generator, the total number operator
\eq \widehat{N}:={\ad}^\al_\i a^\i_\al~. \qe
So the oscillators ${\ad}^\al_\i$ and hence the Fock space $\F^{\mathrm{Total}}$, carry a $su(n+1) \times su(n) \times u(1)$ representation.

We also introduced composite operators $\tA^\al,\,\tA_\al^\dagger$ formed by using the completely antisymmetric tensor, $\epsilon_{\i_1\cdots \i_n}$, to form an $su(n)$ singlet from the oscillators $a^\i_\al$ (and dualizing the $n$ $su(n+1)$ indices):
\eq \tA^\al :=\frac{1}{n!}\epsilon^{\al \bt_1\cdots \bt_n}\epsilon_{\i_1\cdots \i_n} a^{\i_1}_{\bt_1} \cdots a^{\i_n}_{\bt_n}~.\label{compositeAtilde}\qe
$\tA_\al^\dagger$ generate a subspace of $\F^{\mathrm{Total}}$ which we call the {\it reduced Fock space}, $\F$, which is a singlet under $su(n)$. It naturally decomposes into a direct sum of subspaces, $\F_L$, associated with the
eigenvalue, $L$, of the number operator on the reduced Fock space
\eq \widehat\calN :=\frac{1}{n}{\ad}^\al_\i a^\i_\al = \frac{\widehat{N}}{n} \qe so that
\eq \F={\mbox{\scriptsize$\displaystyle{\bigoplus^{\infty}_{L=0}}$}}{\cal F}_L~ \qe
where ${\cal F}_L$ is spanned by $\tA^\dagger_{\alpha_1}\cdots \tA^\dagger_{\alpha_L} |0>$.

The above identifies ${\cal F}$, the subspace of $\F^{\mathrm{Total}}$ which carries the singlet $\ID$ representation of $su(n)$. In  the decomposition of
$\F^{\mathrm{Total}}$ into irreducible representations of the algebra
$su(n+1)\times su(n)\times u(1)$ all irreps., ${\cal R}$, of $u(n)$ appear with
multiplicity one,
\eq \F^{\mathrm{Total}}=\oplus_{\cal R} \F_{\cal R} ~.\qe
For example, if $\cal R$ is the fundamental representation, $\tyng(1)\,$, of $u(n)$ then $\F_{\tyng(1)}=\oplus_{L=0}^\infty\F^L_{\tyng(1)}$ is given by the span of
\eq(\ad)^\bt_\i \F= \bigoplus_{L=0}^\infty(\ad)^\bt_\i\F^L \label{chargenorm}\qe
(we normalise the $U(1)$ charge so that $(a^\dagger)^\bt_\i$has charge $1/n$
and $\tA^\dagger_\al$ charge 1, then states in  $\F^L_{\tyng(1)}$ have charge
$L+\frac 1 n $).

The composite oscillators do not satisfy the Heisenberg algebra although they do generate a closed subspace of $\F^{\mathrm{Total}}$. We have shown \cite{Dolan:2006tx} that it is possible to normalize them in such a way so that they do however satisfy the Heisenberg algebra on the reduced Fock space, $\F$.
Define
\begin{align}
A^\al &:=\tA^\al \sqrt{\frac{\widehat{\cal N}!}{(\widehat{\cal N}+n-1)!}}
\label{compositeA}\\
A_\al^\dagger &:=\sqrt{\frac{\widehat{\cal N}!}{(\widehat{\cal N}+n-1)!}}\tA_\al^\dagger~.\label{compositeAdagger}
\end{align}
Then denoting elements of $\F_L$ by
\eq \ket{\avec}:=\frac{1}{L!}A^\dagger_{\al_1} \cdots A^\dagger_{\al_L} \ket{0} \qe
or
\eq \ket{\widetilde{\avec}}:=\frac{1}{L!}\tA^\dagger_{\al_1} \cdots \tA^\dagger_{\al_L} \ket{0} \qe
we have
\eq [A^\al,\,A^\dagger_\bt]=\dl^\al_\bt \label{A commutator}\qe
on such elements.

It is then easy to see that the standard Heisenberg algebra / Fock space relations hold
\begin{align}
\braket{\avec}{\bvec}&=S^{\avec}_{\bvec}:=\frac{1}{L!}\dl^{\al_1}_{ \{ \bt_1 } \cdots \dl^{\al_L}_{\bt_L \} }\\
\widehat{\cal N}_A\ket{\avec}&:=A^\dagger_\bt A^\bt \ket{\avec}=\widehat{\cal N}\ket{\avec}=L\ket{\avec}~,
\end{align}
though $\widehat{\cal N}_A\ne\widehat{\cal N}$ on the full Fock space
generated by $a^\dagger{}^\al_\i$, they are only equal on the reduced Fock
space.

We can also realise the generators of $SU(n+1)$ in the usual way
\eq \hat{L}_a:=A^\dagger_\al \left(\frac{\lambda_a}{2}\right)^\al{}_\bt A^\bt~. \qe
On the reduced Fock space these can be shown to be equivalent to the $SU(n+1)$ generators $\hat{J}_a :=(\frac{\bar{\lambda}_a}{2})_\al {}^\bt \hat{J}^\al {}_\bt$ written in term of the oscillators $a^\i_\al$ and $(\ad)^\al_\i$
\eq \hat{J}_a \ket{\avec} =\hat{L}_a \ket{\avec}~, \qe
though $\hat{J}_a\ne \hat{L}_a$ on $\F^\mathrm{Total}$.

The $su(n)$ singlet nature of $A^\dagger_\al$ is shown by
\eq
\hat{J}_\i{}^\j \ket{\avec}=\dl_\i^\j \widehat{\cal N} \ket{\avec}=L\dl_\i^\j \ket{\avec}\label{Jijsinglet}\qe
and
\eq (\ad)^\al_\i A^\dagger_\al=0~.\qe

\noindent
The action of an $su(n+1)$ generator on a singlet state is given by
\eq
{\hat J}^\gamma_{~~\beta}\ket{\avec}=L\delta^\gamma_\beta\ket{\avec}-
\frac{1}{\sqrt{L!}} \sum_{k=1}^{L}
\delta_{\alpha_k}^\gamma A^\dagger_\beta A^\dagger_{\alpha_1} \cdots A^\dagger_{\alpha_{k-1}} A^\dagger_{\alpha_{k+1}} \cdots A^\dagger_{\alpha_L}|0>.
\label{numberrel} \qe

$\F_L$ is a left module for the noncommutative algebra of functions, $\calA_L:=\F_L \otimes \F_L^*$ and a basis is $\ket{\avec}\bra{\bvec}$.
Since the dimension of $\F_L$ is
\eq
\mathrm{dim}(\: \hbox{{\tiny$\overbrace{\yng(5)}^L$}}\:)_{su(n+1)}=\frac{(L+n)!}{L!n!}~=:d_n (L),
\qe
$\calA_L$ is isomorphic to $Mat_{d_n(L)}$ and is the space of linear mappings from $\F_L$ to $\F_L$. The dimension can also be seen by taking the trace of the identity matrix,
\eq
Tr (\ID)=Tr(\ket{\avec}\bra{\avec})=\braket{\avec}{\avec}=d_n (L)~.
\qe

More generally we can extend the construction to non-square matrices
and consider $\F_L \otimes \F_{L'}^*$ as the space of linear mappings from $\F_{L'}$ to $\F_L$.  As shown in \cite{Dolan:2006tx}
these describe noncommutative equivariant line bundles over
$\CP^n=SU(n+1)/U(n)$: elements of the set of such maps
are carrier spaces for $SU(n+1)$, invariant under $SU(n)$ and have a $U(1)$ charge proportional to $q:=L-L'$. Such
matrices form a left
$Mat_{d_n (L)}$ module and a right $Mat_{d_n (L')}$ module. We will also consider $\F_L \otimes \F_{\cal R}$, corresponding to equivariant vector
bundles over $\CP^n$, in the next section, but for the remainder of this
section we focus on line bundles $\F_L \otimes \F^*_{L'}$.

A general $d_n(L)\times d_n(L')$ matrix ${\bf M} \in \F_L \otimes \F^*_{L'}$ takes the form
\eq
{\bf M}=M^{\avec}{}_{\bvec} \ket{\avec}\bra{\bvec} \qe
where $M^{\avec}{}_{\bvec}=M^{\al_1\cdots\al_L}_{\bt_1\cdots\bt_{L'}}$ are complex co-efficients.
The action of $su(n+1)$  on $\F_L \otimes \F^*_{L'}$ is $(\hat{L}_a \F_L \otimes \F^*_{L'}) + (\F_L \otimes  \F_{L'}^* (-\hat{L}_a) )$; this leads us to define
\begin{align}
\hat{\calL}_a {\bf M}:=&(\hat{L}_a)^{{\ttL}} {\bf M}- (\hat{L}_a)^{\ttR}{\bf M}, \non\\
=&\hat{L}_a {\bf M}- {\bf M} \hat{L}_a \end{align}
where the superscripts $^{\ttL}$ and $^{\ttR}$ indicate left and right
action respectively,
and the operators $\hat{\mathcal{L}}_a$ can be seen as right invariant vector fields induced from the left action of $SU(n+1)$ on ${\CP}^n$, they satisfy the
$su(n+1)$ commutation relations.

A Laplacian for $\mathcal{F}_L \otimes \mathcal{F}_{L^{\prime}}$ can be immediately constructed from left actions

\eq
\Delta{\bf M} = \hat{\mathcal{L}}_a \hat{\mathcal{L}}_a {\bf M}~.
\qe
This is just the $su(n+1)$ quadratic Casimir operator corresponding to the appropriate representation, explicitly

\eq   \label{lap}
\Delta = (\hat{L}_a \hat{L}_a)^\ttL \otimes {\bf 1} + {\bf 1}\otimes (\hat{L}_a \hat{L}_a)^\ttR -    2  \hat{L}_a^\ttL \otimes \hat{L}_a^\ttR  .
\qe
It was also shown in \cite{Dolan:2006tx} that a diagonal basis for
$\F_L\otimes \F_{L'}$ can be realised as the set of (here un-normalized) polarization tensors which are eigenstates of the Laplacian,
${{\bf D}_{\avec_{l+q}}}^{\bvec_{l}} \in \F_L\otimes \F_{L'}^*$, given by

\eq \label{polariztens}
{{\bf D}_{\avec_{l+q}}}^{\bvec_{l}}=
{{{\cal P}_{\avec_{l+q},\svec_{l}}}}^{\bvec_{l},\tauvec_{l+q}}
|\tauvec_{l+q},\gvec_{L'-l}\rangle\langle\svec_{l},\gvec_{L'-l}|
\qe
where ${{{\cal P}_{\avec_{l+q},\svec_l}}}^{\bvec_l,\tauvec_{l+q}}$ is the projector that removes all traces associated with contractions of the free indices of ${\cal F}_L$ and ${\cal F}_{L'}^*$; $\avec_{l}$ indicates $l$ free indices $\al_1\cdots \al_{l}$.
We write a non-square eigenmatrix of the Laplacian, ${\bf \Phi}_{l_\ttL,l_\ttR}$,
as
\eq{\bf \Phi}_{l_\ttL,l_\ttR} = \phi^{\boldsymbol{\alpha}}_{~~\boldsymbol{\beta}} | \boldsymbol{\alpha}_{l_\ttL},~\boldsymbol{\gamma} \rangle \langle \boldsymbol{\gamma},~\boldsymbol{\beta}_{l_\ttR}|
\qe
where the coefficients $\phi^{\boldsymbol{\alpha}}_{~~\boldsymbol{\beta}}$ are traceless under contractions of any pair of upper and lower indices and here $l_\ttL=l_\ttR+q$.

The spectrum of the Laplacian can be obtained from its action on
${\bf \Phi}_{l_\ttL,l_\ttR}$

\eq \label{lapphi}
\Delta {\bf \Phi}_{l_\ttL,\,l_\ttR} = \frac{1}{2} \Big( l_{\ttL} (l_{\ttR}+n) + l_{\ttR} (l_{\ttL}+n)+\frac{n(l_\ttL-l_\ttR)^2}{n+1}\Big){\bf \Phi}_{l_\ttL,l_\ttR}~.
\qe
Polarization tensors (\ref{polariztens}) form a complete orthogonal basis for $\mathcal{F}_L \otimes \mathcal{F}_{L^{\prime}}$ under the trace inner product:

\eq
\langle {\bf M} , {\bf N} \rangle := Tr({\bf M}^{\dagger} {\bf N}).
\qe

The matrices ${\bf \Phi}_{l_\ttL,l_\ttR}$ represent sections of line bundles,
at the fuzzy level.
Let ${\bL}$ denote the tautological line-bundle over $\CP^n$ in the
exact sequence
\eq 0\rightarrow {\bL}\rightarrow {\bf V} \rightarrow {\bf F} \rightarrow 0 \label{exactsequence}\qe
where ${\bf V}=\CP^n\otimes {\mathbb C}^{n+1}$ is the trivial rank $n+1$ complex
bundle over $\CP^n$ and ${\bf F}$ is the rank $n$ bundle over $\CP^n$ induced by the
above sequence.  Then
${\bL}$ has
first Chern number $C_1({\bL})=-1$ and ${\bf \Phi}_{l_\ttL,l_\ttR}$
is a fuzzy section of ${\bL}^{l_\ttL-l_\ttR}$.
Monopole bundles over fuzzy complex projective spaces were constructed, in terms of projective modules,
in \cite{CarowWatamura:2004ct}.

\subsection{Covariant Derivatives} \label{CovD}
In \cite{Dolan:2006tx} we also introduced operators\footnote{For $\CP^1$ these operators were already given by Grosse {\it et al.} \cite{Grosse:1995jt}.} corresponding to right action of the compliment of the $u(n)$ subalgebra in $su(n+1)$ on equivariant vector bundles over $\CP^n$, $\hat{K}_\i,\,\hat{K}_{\bar{\imath}}$, which act as covariant derivatives on the spaces of noncommutative functions $\calA_L=\F_L\otimes \F^*_L$ and noncommutative line bundles $\F_L  \otimes \F^*_{L-q}$
\eq
\begin{array}{lcllll}
\label{Kis_as_module_maps}
\hat K_{\imath}:=&{(A^\dagger_\alpha)}^{\ttL} {({({a^\dagger})}^\alpha_\imath)}^{\ttR}
&:& {\cal F}_{L}\otimes {\cal F}^*_{L'}
&\longmapsto& {\cal F}_{L+1}\otimes {{\cal F}^*_{L',\imath}}\\
\hat K_{\bar\imath}:=&{(A^\alpha)}^{\ttL} {(a_\alpha^\imath)}^{\ttR}
&:&{\cal F}_{L}\otimes {\cal F}_{L'}^*
&\longmapsto& {\cal F}_{L-1}\otimes {{\cal F}^*_{L',\bar\imath}}~,
\end{array}
\qe
where we have denoted the subspace of Fock space spanned by vectors of the form
\eq
{(a^\dagger)}^{\alpha}_\imath A^\dagger_{\alpha_1}\dots A^\dagger_{\alpha_L}\ket{0}
\qe
by ${\cal F}_{L,\i}={\cal F}_L^{~\bi}$. The remainder of the left invariant vector fields generating $U(n)$ acting on $\CP^n$ are given by $ [\hat{K}_{\i} ,\,\hat{K}_{\bar{\j}}]$. For a matrix ${\bf M}_{{\cal R}} \in \F_L \otimes \F^*_{\cal R}$ we find
\eq
[\hat K_\imath,\,\hat K_{{\bar\jmath}} ] {\bf M}_{\cal R}=-{\bf M}_{\cal R}({{\hat J}_\imath}^{~\jmath}-\delta_\imath^\jmath\hat {\cal N})+2\delta_\imath^\jmath\hat K_0
{\bf M}_{\cal R}~,
\qe
where $\hat K_{0}:=\frac{1}{2}(\hat{\cal N}_A^{\ttL}-\hat{\cal N}^{\ttR})$, and we easily see that $\F_L \otimes \F^*_L$ is invariant under the right action of $U(n)$ as expected.
The Laplacian acting over ${\bf M}_{\mathcal{R}}$ is naturally the quadratic $su(n+1)$ Casimir on $\mathcal{F}_L\otimes \mathcal{F}_{\mathcal{R}}^*$

\eq
\Delta {\bf M}_{\mathcal{R}} = (\hat{L}_a^{\ttL} - \hat{J}_a^{\ttR})^2 {\bf M}_{\mathcal{R}}
\qe
and, when
%
%
$\mathcal{R}$ is the symmetric combination of $L'$  anti-fundamental
representations of $su(n+1)$
$\hbox{{\tiny
${\overbrace{\young(\ \scdots \ ,\weevdots \scdots \weevdots ,\ \scdots \ )}^{L'}}\left.\vline height 12pt depth 7pt width 0pt \right\} n$}}$.
The Laplacian can be constructed in terms of right actions as

\eq
\Delta_K = \frac{1}{2}(\hat{K}_{\i} \hat{K}_{\bar{\i}} + \hat{K}_{\bar{\i}}\hat{K}_{\i} )+ \frac{2n}{n+1}\hat{K}_0^2.
\qe
On eignematrices $\Phi_{l_\ttL,l_\ttR}$, $\Delta_K$ coincides with (\ref{lap}) having the same eigenvectors and spectrum (\ref{lapphi}) as can be verified from the relations

\eq \label{KiKbi on nonsquare}
\hat{K}_{\i}\hat{K}_{\bar{\i}} {\bf \Phi}_{l_\ttL,l_\ttR} = l_{\ttL}(l_{\ttR}+n) {\bf \Phi}_{l_\ttL,l_\ttR}
\qe
and
\eq
\hat{K}_{\bar{\i}}\hat{K}_{\i} {\bf \Phi}_{l_\ttL,l_\ttR} = l_{\ttR}(l_{\ttL}+n) {\bf \Phi}_{l_\ttL,l_\ttR}~,
\qe
with $\hat K_0=(l_\ttL - l_\ttR)/2$.

\section{Spinors}
\label{spinors}
On a $2n$ dimensional real manifold, with a Euclidean metric, the $\Gamma$-matrices can be chosen
to satisfy
\eq\{\Gamma^\mu,\Gamma^\nu\}=-2\delta^{\mu\nu}, \qquad \mu,\nu=1,\ldots 2n
\label{Clifford}\qe
in an orthonormal basis
and to be anti-Hermitian, $(\Gamma^\mu)^\dagger=-\Gamma^\mu$.
On an $n$-complex dimensional K\"ahler manifold, such as $\CP^n$, we can define
$\gamma^\imath =(-i\Gamma^\imath +\Gamma^{\imath +n})/2$, with $\imath=1,\ldots,n$.
The Dirac algebra in this basis is
\begin{align} \{\g^\i,\,\g^\j\}&=\{\g^{\bi},\,\g^{\bj}\}=0 \\
\{\g^\i,\,\g^{\bj} \}&=\dl^{\i \bj} ~,\end{align}
where $\g^\bi= (\g^\i)^\dagger$.
One can think of $\g^{\bi}$ as fermionic creation operators and $\g^\i$ as the Hermitian conjugate fermionic annihilation operators. The spinors, on which the $\gamma$-matrices act, can be constructed by acting on a Clifford vacuum $\ket{\Omega}$ defined by $\g^\i \ket{\Omega}=0$. A general spinor field then has an expansion \cite{Green:1987mn}
\begin{align} \Psi &=\psi_{0}\ket{\Omega}+\psi_{\bi} \g^{\bi} \ket{\Omega}+
\frac{1}{2!}\psi_{\bi \bj}\g^{\bi} \g^{\bj} \ket{\Omega}+\cdots+\psi_{\bi_1 \bi_2 \cdots \bi_n}\g^{\bi_1} \g^{\bi_2} \cdots \g^{\bi_n} \ket{\Omega}~. \label{spinor} \end{align}
\comment{Of course, at the commutative level, we are not working with local holomorphic coordinates but global homogeneous coordinates $z^\al,\,\bar{z}_\al$. However, we
will see that the spinors take the same form as above but where $\psi_{\bi_1 \cdots \bi_k}$ are functions of $z^\al$, $\bar{u}^\i_\al$ and their
conjugates (see Appendices A and B of \cite{Dolan:2006tx}).}
The chirality operator is $\Gamma:=\prod_{\i=1}^n\,[\gamma^\i,\gamma^{\bar\imath}]$
satisfying
$\{\Gamma,\gamma^\i\}=\{\Gamma,\gamma^{\bar\imath}\}=0$ and
$\Gamma^2={\bf 1}$. It has the effect of changing the sign of the terms
in $\Psi$ with an odd number of $\gamma$-matrices.

Because the holonomy group is $U(n)$, we do not, in general, expect $\psi_{0}$ to be neutral under the $U(1)$ part of the $U(n)$-valued
spin connection, i.e. $\psi_0$ is a charged scalar field.

Let $\wedge^{0,\,k}\overline{T^* \CP^n}$ be the bundle of $(0,k)$-forms and
\eq \wedge^{0,\,*}T\CP^n=\bigoplus_{k=0}^n \wedge^{0,\,k}\overline{T^* \CP^n}\qe
be the bundle of linear combinations of $(0,k)$-forms.
Then the spin bundle $S(\CP^n)$ can be defined
by \cite{Lawson:1990}
\begin{align} S(\CP^n):=\wedge^{0,\,*}T\CP^n \otimes \bL^{\frac{n+1}{2}}~.
\label{diracspinor}\end{align}
It only exists for odd $n$ as can be seen from the power of the line bundle $\bL$, reflecting the fact that only $\CP^n$ with odd $n$ has spin structure.
It is possible to overcome the odd $n$ restriction by considering spin$^c$ structures. We simply tensor with different integer powers of $\bL$:
\eq S_q(\CP^n):= \wedge^{0,\,*}T\CP^n \otimes \bL^{q}~, \quad q\in \mathbb{Z}.
\label{chargeq}\qe
These bundles exist for all $n$ and give general spin$^c$ bundles --- the
case $q=0$, where $\psi_0$ in (\ref{spinor}) has no $U(1)$ charge
and so is just a function, is canonical spin$^c$. Sections of $S_q(\CP^n)$
can be viewed as
linear combinations of $(0,k)$-forms, each with charge $q$, and we shall
denote the space of charged $(0,k)$-forms by $\Lambda^{0,\,k} (\CP^n,\,q)$.
Note that for canonical spin$^c$, $S_0(\CP^n)$ is just the bundle of linear combinations of
$(0,k)$-forms, with $q=0$, while ordinary spinors,
{\it i.e.} sections of $S(\CP^n)$ in (\ref{diracspinor}) when
$n$ is odd, can be identified with
linear combinations of $(0,k)$-forms carrying a $U(1)$ charge $(n+1)/2$.
For $SU(n)$ holonomy such complications do not occur and ordinary
spinors {\it are} neutral $(0,k)$-forms.

The Dirac operator
\eq \Dirac=\g^\bi D_\bi + \g^\i D_\i =\bar{\cal D} + \bar{\cal D}^\dagger,\qe
where $\bar{\cal D}:=\gamma^{\bi} D_{\bi}$ and $\bar{\cal D}^\dagger:=
\gamma^\i D_\i$ its adjoint,
acts on sections of $S_q(\CP^n)$ and can be constructed once
the relevant  co-variant
derivatives, $D_\bi$ and $D_\i$, are known. The bundles (\ref{chargeq})
are holomorphic in the sense that
\eq F_{\i\j}=[D_\i,D_\j]=0, \qquad F_{\bi\bj}=[D_\bi,D_\bj]=0.\qe

\bigskip

\section[Noncommutative Spinors]{Noncommutative Spinors\footnote{We shall call elements of the projective modules corresponding to sections of the spin$^{\mathrm{c}}$ bundle(s) over fuzzy $\CP^n$ ``noncommutative spinors'', ignoring the question of a wedge product for their charged $(0,k)$ form constituents. We will not give a wedge product but work only with modules.}}
\label{noncommutative spinors}

In section \ref{CovD} we defined covariant derivative operators, $K_\i$ and
$K_\bi$, acting on fuzzy $\CP^n$ and we shall now show that these
are indeed the correct  objects to use in the fuzzy Dirac equation.
We propose
\eq
\Dirac=\bar{\D}+\bar{\D}^\dagger := \g^\bi \K_\bi+\g^\i\K_\i
\label{DiracK}
\qe
as a noncommutative massless Dirac operator.
We shall show that this has the same spectrum as the known continuum cases
cut-off at a finite level.
The Dirac operator (\ref{DiracK})
clearly anti-commutes with the chirality operator, $\{\Dirac,\Gamma\}=0$.

A detailed discussion of the complete spectrum of (\ref{DiracK}) with an
explicit construction
of all eigenspinors together with their corresponding eigenvalues and degeneracies is given in two appendices, where it is shown that
the non-zero eigenvalues $\lambda$ are
labelled by three integers $q$, $l$ and $k$,
with $l=0,\ldots,L$ and
$k=0,\ldots,n-1$ constrained by $l+q\ge k+1$, and are given by
\eq \lambda=\pm\sqrt{(l+q)(l-k+n)},\label{spinoreigenvalues}\qe
with degeneracies
\eq g_\lambda =\frac{(l+n)!(l+q-k-1+n)!(2l+q-k+n)}{l!k!n!(l+q-k-1)!(n-k-1)!(l+n-k)(l+q)}~\label{spinordegeneracies}\qe

For the moment, for simplicity, we shall motivate the choice (\ref{DiracK})
with a discussion of
the zero modes, explicitly constructing all such modes for any $n$
and proving that everything is consistent with known results.

The noncommutative version of sections of the
line bundles $\bL^{q}$, $q\geqslant0$
have already been identified: they are just the bi-modules
\eq \F_{L'+q} \otimes \F^*_{L'} \qe
and have $U(1)$ charge $q$ as measured by $2\K_0$ (in the
conventions adopted here the charge is the opposite of the first Chern number
$C_1(\bL^q)=-q$). Similarly the noncommutative counterpart of sections
of the line bundles $\bL^q$ with $q<0$
are described by the bi-modules
\eq  \F_L \otimes \F^*_{L+|q|}~.\qe
Specific elements of these line-bundles are zero-modes of the Dirac operator.

The Atiyah-Singer index, $\nu$,
for zero modes of the Dirac operator on sections of
the bundle $S_q(\CP^n)\approx S_0(\CP^n)\otimes \bL^q$, where
$q\in {\bf Z}$, is \cite{Dolan:2002ck}
\eq\nu=\frac{(1-q)\cdots (n-q)}{n!}.
\label{Lindex}\qe
For $q\le 0$ this
is the same as the dimension of
the symmetric representation of $su(n+1)$ with a Young tableau consisting
of a row of $|q|$ fundamental representations:
\eq
\overbrace{\yng(5)}^{|q|}\;.\label{nuqplus}\qe
For $q\ge (n+1)$ the magnitude $|\nu|$ of the index (\ref{Lindex})
is the same as the dimension of the representation
consisting of $q-n-1$ symmetrised anti-fundamentals of $su(n+1)$,
\eq
\overbrace{\young(\ \scdots \ ,\weevdots \scdots \weevdots ,\ \scdots \ )}^{q-n-1}\left.\vline height 17pt depth 17pt width 0pt \right\} n\;.
\label{nuqminus}
\qe
For $0<q<n+1$ the index is zero. For odd $n$, $\nu$ is positive for
$q\le 0$ and negative for $q\ge n+1$ while, for even $n$, $\nu$ is always
positive.

These Young tableaux reflect the symmetries of the wave-functions corresponding to the
zero-modes.
Our proposal for the Dirac operator is
\eq
\Dirac=(A^\alpha)^\ttL (a^\i_\alpha)^\ttR \gamma^{\bar\imath}
+ (A^\dagger_\alpha)^\ttL({a^\dagger} {}_\i^\alpha)^\ttR \gamma^\i.\label{UoneDirac}
\qe
Zero modes for $q\le 0$ are
\eq \bm\psi_q^{\vectorindex{\alpha}}=  |0><0|A^{\alpha_1}\cdots A^{\alpha_{|q|}} \otimes |\Omega>,\label{spinq}\qe
while the zero modes for $q\ge n+1$ are
\eq \widetilde{\bm\psi}_{q,\vectorindex{\alpha}}
= A^\dagger_{\alpha_1}\cdots A^\dagger_{\alpha_{q-n-1}} |0><0| \otimes
\gamma^{\bar 1}\cdots \gamma^{\bar n}|\Omega>.\label{spinqbar}\qe

These are the smallest possible expressions of the zero modes for a given
$q$ --- they can be embedded into bigger matrices using
\eq|\vectorindex{\mu}_L><\vectorindex{\mu}_L|A^{\alpha_1}\cdots A^{\alpha_{|q|}} \otimes |\Omega>\label{spinnq}\qe
and \eq A^\dagger_{\alpha_1}\cdots A^\dagger_{\alpha_{q-n-1}} |\vectorindex{\mu}_L><\vectorindex{\mu}_L| \otimes
\gamma^{\bar 1}\cdots \gamma^{\bar n}|\Omega>\label{spinpq}\qe
respectively.

Given that we have assigned the $U(1)$ charges
$Q(A^\dagger_\alpha)=1$ to $A^\dagger_\alpha$, and $Q(a^\dagger{}^\alpha_\i)=1/n$
to $a^\dagger{}^\alpha_\i$,
demanding that the Dirac operator commutes with the charge
operator requires assigning the charge $Q(\gamma^{\bar\imath})=(n+1)/n$
to the $\gamma$-matrices.
The vacuum must then be charged, $Q(|\Omega>)=v$, and following the
arguments in \cite{Green:1987mn},
$Q(\gamma^{\bar 1}\cdots \gamma^{\bar n}|\Omega>)=n+1 +v=-v$
so $Q(|\Omega>)=-(n+1)/2$ in the conventions used here.
The $U(1)$ charge of the states in (\ref{spinnq}) and (\ref{spinpq})
is then $q-\frac{n+1}{2}$, consistent with the statement that Dirac spinors
(with $q=(n+1)/2$ for odd $n$) have zero charge.
These charge assignments can be obtained from the charge operator
\eq \widehat Q=(A^\dagger_\alpha A^\alpha )^{\tt L} -\frac{1}{n}
\left( (a^\dagger)_\alpha^\i a^\alpha_\i \right)^{\tt R}
-\frac{n+1}{2n} [\gamma^\i, \gamma^\bi]
\qe
acting on the states (\ref{spinnq}) and (\ref{spinpq}) above and
(\ref{fullfuzzyspinor}) below.\footnote{Note that
$q\ne L'-L$ in (\ref{spinpq}), because the $\gamma$-matrices carry
charge.  From here on we drop the distinction between $L$ and $L'$
adopted up till now  --- the size of the matrices should be clear from
the context.}  In all cases the total charge is $Q=q-\frac{n+1}{2}$,
the last term being the contribution from the Clifford vacuum $|\Omega>$.

In terms of coherent state representations the wave-functions (\ref{spinq})
and (\ref{spinqbar})
are symmetric monomials, of order $|q|$ in
$z^\alpha$ and of order $q-n-1$ in $\bar z_\alpha$ respectively.
These form a basis for all zero-modes of equivariant line bundles.  
The family of zero modes (\ref{spinq}) and (\ref{spinqbar})
transform as irreps. of $SU(n+1)$, 
(\ref{nuqplus}) and (\ref{nuqminus}) respectively, whose dimensions
agree with the index (\ref{Lindex}), up to a sign.  For a given equivariant
line bundle all zero modes are of the same chirality.
In the coherent state representation, the most general zero-mode is
a linear superposition of the respective monomials in homogeneous
co-ordinates $z^\alpha$ and $\bar z_\alpha$.

This provides strong support for the form (\ref{DiracK}) for
a \lq\lq universal'' Dirac operator acting
on all $S_q(\CP^n)$.  A general spinor on fuzzy $\CP^n$ requires
replacing the co-efficients $\psi_{\bi_1\cdots\bi_k}$ in (\ref{spinor}) with
Fock space operators,
\eq
\Psi =\sum_{k=0}^n \frac{1}{k!}{\bm\psi}_{\bi_1\cdots\bi_k}
\g^{\bi_1} \cdots \g^{\bi_k} \ket{\Omega}\label{fullfuzzyspinor}\qe
where ${\bm\psi}_{\bi_1\cdots\bi_k}$ are linear combinations of the operators
\eq {\bm\psi}^k_{\avec}{}^{\bvec}=A^\dagger_{\alpha_1}\cdots A^\dagger_{\alpha_{L+q-k}}|0><0|A^{\beta_1} \cdots A^{\beta_L}a^{[\i_1}_{\alpha_{L+q-k+1}}\cdots a^{\i_k]}_{\alpha_{L+q}}.
\label{fuzzyspinorcomponents}\qe
A full discussion of all eigenstates, non-zero
eigenvalues (\ref{spinoreigenvalues})
and the construction of the corresponding eigenstates is somewhat technical
and is left to an appendix.

\section{Charge Conjugation}

First recall standard results in ordinary commutative continuum space,
{\it e.g.} \cite{Kugo:1982bn}.
The charge conjugation matrix is required to satisfy
\eq (\Gamma^\mu)^T=\pm C \Gamma^\mu C^{-1},\qquad \mu=1,\ldots,2n
\label{Cmatrix}\qe
where, in $2n$ dimensions, either sign is possible for a given $n$.
Write
the usual continuum Dirac operator in flat space as
\eq i\partial \kern -6pt \slash = i\Gamma^\mu\partial_\mu.\qe
We choose signature
$(-,\ldots,-)$ and use the Clifford algebra (\ref{Clifford}).

In our conventions $(\Gamma^\mu)^\dagger = -\Gamma^\mu$
are anti-Hermitian, $\gamma^\i=(-i\Gamma^\i + \Gamma^{\i+n})/2$ and
the flat space Dirac operator is, using complex co-ordinates,
\eq
i\partial \kern -6pt \slash=\gamma^\i\partial_\i + \gamma^{\bar\imath}\partial_{\bar\imath}.\qe
For example for $n=1$ we can take $\Gamma^1=i\sigma_1$ and $\Gamma^2=i\sigma_2$,with $\sigma_1$ and $\sigma_2$ Pauli matrices, and then $\gamma^\i=\sigma_+$
is real.

We use the same conventions in curved space, with $\Gamma^\mu$
satisfying (\ref{Clifford}) in an orthonormal basis.  In general we can choose
$\Gamma^\i$ to be symmetric and $\Gamma^{\i+n}$ to be anti-symmetric so
$\gamma^\i$ are real, with
\eq (\gamma^\i)^\dagger = \gamma^{\bar\i}=(\gamma^\i)^T\qquad
\Rightarrow\qquad C \gamma^\i C^{-1}=\pm\gamma^{\bar\i},\label{Camb}\qe
and the charge conjugation matrix must satisfy
\eq [C,\Gamma^i]=0, \qquad \{C,\Gamma^{i+n}\}=0\qe
for the upper sign and
\eq \{C,\Gamma^i\}=0, \qquad [C,\Gamma^{i+n}]=0\qe
for the lower sign.
So $C$ can be chosen to be
\eq C=C_+:=\left\{
\begin{array}{llc}
(\gamma^{\bar 1} +\gamma^1)\cdots (\gamma^{\bar n} +\gamma^n)&=\Gamma^1\cdots\Gamma^n,&
n \ \hbox{odd}  \\
\ & \\
(\gamma^{\bar 1} -\gamma^1)\cdots (\gamma^{\bar n} -\gamma^n)&=
\Gamma^{1+n}\cdots\Gamma^{2n},& n \ \hbox{even}\\
\end{array}\right.\qe
for the upper sign in (\ref{Camb}) and
\eq C=C_-:=\left\{
\begin{array}{llc}
(\gamma^{\bar 1} -\gamma^1)\cdots (\gamma^{\bar n} -\gamma^n)&=-\Gamma^{1+n}\cdots\Gamma^{2n},& n \ \hbox{odd}\\
\ & \\
(\gamma^{\bar 1} +\gamma^1)\cdots (\gamma^{\bar n} +\gamma^n)&=\Gamma^1\cdots\Gamma^n, &
n \ \hbox{even}  \\
\end{array}\right.\qe
for the lower sign.
In general
\eq C_\pm=\bigl(\gamma^{\bar 1} \mp(-1)^n\gamma^1\bigr)\cdots \bigl(\gamma^{\bar n} \mp(-1)^n\gamma^n\bigr).\label{Cmatrixdef}\qe
These satisfy $C_\pm^\dagger C_\pm=1$, $C_\pm^T=(-1)^{n(n\mp 1)/2}C_\pm$
and $C_\pm\Gamma = (-1)^n\Gamma C_\pm$.

In the basis we are using $\gamma^{\bar\i}$ are real and
$|\Omega>$ can be chosen to be real, $|\Omega>=(|\Omega>^*)$,
so
\begin{eqnarray}
C_\pm(\gamma^{\bar\imath_1}\cdots \gamma^{\bar\imath_k}|\Omega>)
&=&(C_\pm\gamma^{\bar\imath_1}C_\pm^{-1})\cdots (C_\pm\gamma^{\bar\imath_k}C_\pm^{-1})C_\pm|\Omega>)\non \\
&=&(\pm 1)^k\gamma^{i_1}\cdots \gamma^{i_k} C_\pm|\Omega>.
\end{eqnarray}
It follows from (\ref{Cmatrixdef}) that
\eq C_\pm|\Omega>=\gamma^{\bar 1}\cdots \gamma^{\bar n}|\Omega>\qe
and so
\begin{eqnarray}
C_\pm(\gamma^{\bar\imath_1}\cdots \gamma^{\bar\imath_k}|\Omega>)
&=&(\pm 1)^k\gamma^{i_1}\cdots \gamma^{i_k}\gamma^{\bar 1}\cdots \gamma^{\bar n} |\Omega>\non \\
&=&(\pm 1)^k\frac{(-1)^{k(k-1)/2}}{ (n-k)!}\epsilon^{i_1\cdots i_n}\gamma^{\bar\imath_{k+1}}\cdots\gamma^{\bar\imath_{n}}|\Omega> \\
&=&\frac{(-1)^{k(k\mp1)/2}}{ (n-k)!}\epsilon^{i_1\cdots i_n}\gamma^{\bar\imath_{k+1}}\cdots\gamma^{\bar\imath_{n}}|\Omega>. \non
\end{eqnarray}

Allowing for the fact that charge conjugation complex conjugates the
coefficients, the final expression for the charge conjugate of a spinor,
\eq \Psi = \bigl(\psi + \psi_{\bar\imath}\gamma^{\bar\imath}+\cdots
+\frac 1 {k!}\psi_{\bar\imath_1\cdots\bar\imath_k}\gamma^{\bar\imath_1}\cdots \gamma^{\bar\imath_k}+\cdots +
\frac{1}{n!}\psi_{\bar\imath_1\cdots\bar\imath_n}\gamma^{\bar\imath_1}\cdots \gamma^{\bar\imath_n}\bigr)|\Omega>,\qe
is
\eq \Psi_c:={\cal C}(\Psi)=\sum_{k=0}^n
\frac{(-1)^{k(k \mp 1)/2}}{ k!(n-k)!}\epsilon^{i_1\cdots i_n}
(\psi_{\bar\imath_1\cdots \bar\imath_k})^*
\gamma^{\bar\imath_{k+1}}\cdots\gamma^{\bar\imath_{n}}|\Omega>.\label{Cdef}\qe

Charge conjugation maps between bundles:
\eq {\cal C}:S_q(\CP^n)\rightarrow S_{-q+n+1}(\CP^n).\qe


In the fuzzy case we can still use (\ref{Cdef}) as the definition of charge
conjugation acting on a spinor but the co-efficients
$\psi_{\bar\imath_1\cdots \bar\imath_k}$ are replaced by linear combinations of
the Fock space operators
\eq
A^\dagger_{\alpha_1}\cdots A^\dagger_{\alpha_{L+q-k}}|0><0|A^{\beta_1} \cdots A^{\beta_L}a^{[\i_1}_{\alpha_{L+q-k+1}}\cdots a^{\i_k]}_{\alpha_{L+q}} \non\qe
and
$\epsilon^{\i_1\cdots \i_n}(\psi_{\bar\imath_1\cdots \bar\imath_k})^*$ by
linear combinations of
\begin{multline}
\epsilon^{\i_i\cdots \i_n}\left(A^\dagger_{\alpha_1}\cdots
A^\dagger_{\alpha_{L+q-k}}|0><0|A^{\beta_1} \cdots
A^{\beta_L}a^{\i_1}_{\alpha_{L+q-k+1}}\cdots
a^{\i_k}_{\alpha_{L+q}}\right)^\dagger \non \\
=\epsilon^{\i_i\cdots \i_n}(a^\dagger)_{\i_1}^{\alpha_{L+q-k+1}}
\cdots(a^\dagger)_{\i_k}^{\alpha_{L+q}}A^\dagger_{\beta_1} \cdots
A^\dagger_{\beta_L}|0><0|A^{\alpha_1}\cdots A^{\alpha_{L+q-k}}.
\non
\end{multline}

\section{Conclusions}
\label{conclusions}
We have extended the previous construction of noncommutative equivariant vector bundles over fuzzy $\CP^n$ in \cite{Dolan:2006tx} to noncommutative spin and
spin$^\mathrm{c}$ bundles $\calS_q(\CP^n)$ (the former only exist for odd $n$),
where $q$ is a power of the tautological
line bundle ${\bf L}$ in (\ref{exactsequence}) --- physically
it represents a background monopole to which the spinor couples with charge $q$.
In the continuum a general spinor (\ref{spinor}) can be decomposed into a sum of irreducible
representations of the $SU(n)$ part of the holonomy group $U(n)$.
Only the totally anti-symmetric
irreps of $SU(n)$, labelled by an index $k=0,\cdots,n$, occur.   The trivial representation
appears twice, for $k=0$ and $k=n$, with different $U(1)$ charges.
Each of these $U(n)$
irreps itself has a harmonic
expansion in terms of irreducible representations of the isometry group $SU(n+1)$
and these determine the eigenvalues (\ref{spinoreigenvalues})
and degeneracies (\ref{spinordegeneracies}) of the Dirac operator.
Although the different $SU(n)$ irreps have different $U(1)$
charges the total $U(1)$ charge on a spinor is the same for all components,
because the $\gamma$-matrices in (\ref{spinor}) themselves carry charge.
The total charge is $Q=q-\frac{n+1}{2}$ where the term $-\frac{n+1}{2}$
is gravitational, {\it i.e.} it
comes from the $U(1)$ part of the Riemann curvature.

In the fuzzy case a general spinor
has the form given in (\ref{fullfuzzyspinor}) and
(\ref{fuzzyspinorcomponents}).
The components are non-square matrices representing operators
on a Fock space which is generated by $n(n+1)$
creation and annihilation operators,
$(a^\dagger)^\al_i$ and $a^\i_\al$, carrying the fundamental representation
of $su(n+1)\times u(n)$, with $U(1)$ charge $+1/n$ and $-1/n$ respectively.
For the $k=n$ and $k=0$ components of the spinor (the $su(n)$
singlets) only composite operators, $A^\dagger_\alpha$ and $A^\alpha$
in (\ref{compositeA}) and (\ref{compositeAdagger}), are required.

One of the more elegant ingredients in the construction is the appearance
of a universal Dirac operator (\ref{DiracK}), with $K_\i$ and $K_\bi$
given in (\ref{Kis_as_module_maps}), which generalises the $\CP^1$ case
developed in \cite{Grosse:1994ed}.
This universal Dirac operator has the same form on all the bundles
$\calS_q(\CP^n)$, the construction automatically takes care of the
different connections for the the different bundles.

\comment{Although the different components of the spinor correspond to matrices
of different sizes we show, at the end of appendix A, that a general
spinor can be promoted to a left-module by embedding the metric algebra ${\cal A}_{L+q-n}$,
which is associated with the $k=n$ component of the spinor, sequentially into
larger matrices, ${\cal A}_{L+q-k}$, up to $k=0$.  This embedding provides the larger dimensional
matrix representations required to act on the spinor on the left.
}
Having constructed fuzzy spinors in a consistent manner we hope in the
future to investigate the possibility of coupling
them to other fuzzy fields, through Yukawa and gauge couplings,
with a view to performing numerical simulations with a
finite number of degrees of freedom, as suggested in \cite{Grosse:1995ar}.

\newpage
\appendix
\section{Eigenspinors of the Dirac operator}

In this appendix we determine the eigenspinors of the operator (\ref{DiracK}).
The continuum $SU(n+1)$ representation theory of exact, co-exact and harmonic
$\bL^{\frac{n+1}{2}}$-valued $(0,k)$-forms on $\CP^n$, for $n$ odd, is known \cite{Cahen89:spect_dirac,Cahen94:errat,Bar96:harmon_spinor}. This imposes very strong restrictions on the structure of modules corresponding to these forms. The modules which are described here
show that our choice of Dirac operator gives the correct spectrum, with just a cut-off at the noncommutative level given by $L$, for the continuum cases known to us: spin structure for odd $n$ \cite{Cahen89:spect_dirac,Cahen94:errat,Bar96:harmon_spinor,SeiSemm:cpn} and spin$^c$ structures for $\CP^2$ \cite{Pope:1980ub,Grosse:1999ci}.

The modules discussed in the text, associated with the line bundles $\bL^q$,
are carrier spaces for $su(n+1)$ representations and decompose into irreducibles as\footnote{For convenience we represent anti-fundamental representations
on young tableaux by an over-bar, thus
\hbox{\tiny $\overline{\yng(1)}:=n\left\{\young(\ ,\weevdots,\ ) \right.$}.
Hence, for example, \hbox{\tiny $ \overbrace{\overline{\young(\ \scdots\ )}}^L=n\overbrace{\left\{
\young(\ \scdots \ ,\weevdots \scdots  \weevdots ,\ \scdots \ )\right.}^L$}.}
\begin{align}
\overbrace{\overline{\yng(4)}}^L \otimes \overbrace{\yng(4)}^{L+q}=& \sum_{l=0}^{L} \overbrace{\overline{\yng(4)}}^l \hspace{-0.8mm}\overbrace{\yng(4)}^{l+q}
\qquad\hbox{for}\quad q\ge 0\\
\overbrace{\overline{\yng(4)}}^{L+|q|} \otimes \overbrace{\yng(4)}^{L} =&
\sum_{l=0}^{L} \overbrace{\overline{\yng(4)}}^{l+|q|} \hspace{-0.8mm}\overbrace{\yng(4)}^{l}\qquad\hbox{for}\quad q\le 0 \label{negq:eq}
\end{align}
respectively.
We shall prefer to use a convention consistent with all values of $q$. Therefore we shall always write modules corresponding to $\bL^q$ as \eq \F_{L+q} \otimes \F^*_{L} \qe and write their representation content as
\eq  \overbrace{\overline{\yng(4)}}^L \otimes \overbrace{\yng(4)}^{L+q}= \sum_{l=0;\;l+q\geqslant0}^{L} \overbrace{\overline{\yng(4)}}^l \hspace{-0.8mm}\overbrace{\yng(4)}^{l+q} ~.\label{su(n+1)content of 0-forms}\qe
Here the sum over $l$ is from $0$ to $L$ but with the constraint that $l+q\geqslant 0$.
For $q<0$ this differs from (\ref{negq:eq}) above only by a change in the cut-off $L$.

We can now write down the space of $\bar{\D}$-exact noncommutative $(0,1)$-forms as the left module
\eq\bar{\D}(\F_{L+q}\otimes \F^*_L\otimes \ket{\Omega})\subset \F_{L+q-1}\otimes \F^*_{L,\bi}\otimes \g^\bi \ket{\Omega}~.\qe
This space has basis elements
\eq \ket{\{\avec}\bra{\bvec}a^\i_{\al_{L+q}\}} \otimes \g^\bi \ket{\Omega} \qe
with $\al_l,\: l=1\ldots L+q$, symmetrised and decomposes as
\eq \sum_{l=0;\;l+q\geqslant 1}^L \overbrace{\overline{\yng(4)}}^l \hspace{-0.8mm}\overbrace{\young(\ \ \ \i)}^{l+q}~,\qe
where {\tiny $\young(\i)$} represents the oscillator $a^\i_\al$ acting on the right as opposed to $\Ad_\al$ acting on the left.
A natural generalisation presents itself: we expect the space of all noncommutative $(0,1)$-forms $\bm{\Lambda}^{0,1}(\CP^n,q)$ to be
given by the module
\eq \bm{\Lambda}^{0,1}(\CP^n,q):=\F_{L+q-1}\otimes \F^*_{L,\bi}\otimes \g^\bi \ket{\Omega}= \mathrm{span} \left( \ket{\avec}\bra{\bvec}a^\i_{\al_{L+q}} \otimes \g^\bi \ket{\Omega}\right) ~. \qe
Its representation theory content is
\begin{align}
\sum_{l=0;\;l+q\geqslant 1}^L \overbrace{\overline{\yng(4)}}^l \hspace{-0.8mm}\overbrace{\young(\ \ \ \i)}^{l+q}\quad \oplus \sum_{l=0;\;l+q\geqslant 2}^L \raisebox{2.4mm}{$\overbrace{\overline{\yng(4)}}^l$} \hspace{-0.8mm}\overbrace{\young(\ \ \ \ ,\i)}^{l+q-1}~,
\end{align}
where the latter term corresponds to the space of harmonic and $\bar{\D}^\dagger$-exact $(0,1)$-forms, which we shall refer to as $\bar{\D}$-nonexact forms\footnote{We shall
see presently that harmonic forms (which are zero modes of ${\kern 2pt \raise 1pt\hbox{$\slash$} \kern -7pt D}$)
exist only for $k=0$ or $n$.}. It is spanned by
\eq \Ad_{[\al_{L+q-1}}\ket{\avec}\bra{\bvec}a^\i_{\al_{L+q}]} \otimes \g^\bi \ket{\Omega}~, \qe
and where in the usual interpretation of Young tableaux, $\al_1\cdots \al_{L+q-1}$ are symmetrised first and then $\al_{L+q-1}$ and $\al_{L+q}$ are anti-symmetrised.

We can write this sum as a tensor product
\eq
\overbrace{\overline{\yng(4)}}^L \hspace{-.8mm} \young(\i)\otimes \overbrace{\yng(4)}^{L+q-1}
\qe
or, on realizing $A^\al a^\i_\al=0$ in terms of Young tableaux as the rule {\tiny $n+1\left\{ \young(\ ,\ ,\ ,\i)\right.=0$}, we can write it as
\eq
\overbrace{\overline{\yng(4)}}^L \otimes \overbrace{\yng(4)}^{L+q-1} \otimes\; \young(\i) ~.\qe

We can continue in this fashion to write down all modules. Acting with $\bar{\D}$ on the space of nonexact $(0,1)$-forms, we obtain exact $(0,2)$-forms. Then, as before, we can write the bases of all noncommutative $(0,2)$-forms,
exact and non-exact, as
\eq \bm{\Lambda}^{0,2}(\CP^n,q):=\F_{L+q-2}\otimes \F^*_{L,\bi_1\bi_2}\otimes \g^{\bi_1}\g^{\bi_2} \ket{\Omega}~,\qe
as
\begin{align}
\Ad_{\al_1}\cdots \Ad_{\al_{L+q-2}}\ket{0}&\bra{0}A^{\bt_1}\cdots A^{\bt_L} a^{\i_1}_{\al_{L+q-1}}a^{\i_2}_{\al_{L+q}} \otimes \g^{\bi_1}\g^{\bi_2} \ket{\Omega} & \hbox{all}\non\\
\Ad_{\{\al_1}\cdots \Ad_{\al_{L+q-2}}\ket{0}&\bra{0}A^{\bt_1}\cdots A^{\bt_L} a^{\i_1}_{\al_{L+q-1}\}}a^{\i_2}_{\al_{L+q}} \otimes \g^{\bi_1}\g^{\bi_2} \ket{\Omega}&\hbox{exact}\non\\
\Ad_{\al_1}\cdots \Ad_{[\al_{L+q-2}}\ket{0}&\bra{0}A^{\bt_1}\cdots A^{\bt_L} a^{\i_1}_{\al_{L+q-1}}a^{\i_2}_{\al_{L+q}]} \otimes \g^{\bi_1}\g^{\bi_2} \ket{\Omega} &\hbox{non-exact}
\end{align}
and their representation content is
\begin{align}
\raisebox{2.3mm}{$\overbrace{\overline{\yng(4)}}^L$}\hspace{-0.1mm}\young(\ione,\itwo) \otimes \overbrace{\yng(4)}^{L+q-2}&
\\
=\sum_{l=0;\;l+q\geqslant 2}^L \raisebox{2.3mm}{$\overbrace{\overline{\yng(4)}}^l$} \hspace{-0.8mm}\overbrace{\young(\ione \ \ \ ,\itwo)}^{l+q-1}&\quad\oplus
\sum_{l=0;\;l+q\geqslant 3}^L \raisebox{4.7mm}{$\overbrace{\overline{\yng(4)}}^l$} \hspace{-0.8mm}\overbrace{\young(\ \ \ \ ,\ione,\itwo)}^{l+q-2}\non
\end{align}
where the first sum on the right hand side represents exact forms
and the second non-exact forms.

Repeating the procedure, we can write down the modules describing noncommutative $\bL^q$-valued $(0,k)$-forms as
\eq \bm{\Lambda}^{0,k}(\CP^n,q):=\F_{L+q-k} \otimes \F^*_{L,\bi_1 \cdots \bi_k} \otimes \g^{\bi_1}\cdots\g^{\bi_k}\ket{\Omega}~.\qe
For $k=1, \ldots, n-1$, the representation content breaks into exact and non-exact forms as
\begin{align}
&\qquad\qquad\raisebox{7mm}{$\overbrace{\overline{\yng(4)}}^L$}\hspace{-0.1mm}\young(\ione,\itwo,\cdot,\ik) \otimes \overbrace{\yng(4)}^{L+q-k}\non\\&= \sum_{l=0;\;l+q\geqslant k}^L \raisebox{7mm}{$\overbrace{\overline{\yng(4)}}^l$} \hspace{-0.8mm}\overbrace{\young(\ione\ \ \ ,\itwo,\cdot,\ik)}^{l+q-k+1} \oplus \sum_{l=0;\;l+q\geqslant k+1}^L \raisebox{9.4mm}{$\overbrace{\overline{\yng(4)}}^l$} \hspace{-0.808mm}\overbrace{\young(\ \ \ \ ,\ione,\itwo,\cdot,\ik)}^{l+q-k}\label{representation content of k-forms}
\end{align}
respectively. The right-hand side of equation (\ref{representation content of k-forms}) is valid for $k=1 \ldots n-1$.

For $k=0$ the representation content is that of (\ref{su(n+1)content of 0-forms}).  This contains zero modes of the Dirac operator for $q\le 0$, as given in
(\ref{spinnq}), namely
\eq \ket{\vv{\mu}_L}\bra{\vv{\mu}_L \avec_{|q|}} \otimes \ket{\Omega}~,\qquad \qquad \overbrace{\mbox{\scriptsize $\overline{\yng(4)}$}}^{|q|},\label{zeromodesk=1}\qe
which are annihilated by $\Dirac$ since $\K_{\bi} \ID=0$.

For $k=n$ the decomposition of $\F_{L+q-n}\otimes \F^*_{L+1}$ is
\eq
\overbrace{\overline{\yng(4)}}^{L+1} \otimes \overbrace{\yng(4)}^{L+q-n}=\sum_{l=0;\;l+q\geqslant n+1}^{L+1} \overbrace{\overline{\yng(4)}}^l \hspace{-0.808mm}\overbrace{\yng(4)}^{l+q-n-1} ~,\qe
which contains zero modes, as given in (\ref{spinpq}), when $q\geqslant n+1$
\eq
\ket{\avec_{q-n-1}\vv{\mu}_{L+1}}\bra{\vv{\mu}_{L+1}} \otimes \g^{\bar{1}}\cdots\g^{\bar{n}}\ket{\Omega}~,\qquad \qquad \overbrace{\mbox{\scriptsize $\yng(4)$}}^{q-n-1}~.\label{zeromodesk=N}\qe
They are zero modes of the Dirac operator since $\K_\i \ID=0$. There are no zero modes for $0<q<n+1$.

Therefore the tensor product (\ref{representation content of k-forms}) contains just the representations which are carried by eigenoperators of $\Dirac^2$ that have non-zero eigenvalues.
\\
For completeness we give the dimension of the irreducible representations given above:
\begin{align}
\mathrm{dim}\quad & k+1 \left\{\vphantom{\young(\ \ \ \ ,\cdot ,\cdot,\ )} \right. \raisebox{7mm}{$\overbrace{\overline{\yng(4)}}^l$} \hspace{-0.808mm}\overbrace{\young(\ \ \ \ ,\cdot ,\cdot,\ )}^{l+q-k}\non \\
=&\frac{(l+n)!(l+q-k-1+n)!(2l+q-k+n)}{l!k!n!(l+q-k-1)!(n-k-1)!(l+n-k)(l+q)}~.\label{spinordimensions}
\end{align}
Therefore $\bm{\Lambda}^{0,k}(\CP^n,q)$ has dimension
\eq \mathrm{dim}\left(\bm{\Lambda}^{0,k}(\CP^n,q)\right)= \frac{(L+n+1)!}{L!k!(n-k)!(L+n-k+1)} \frac{(L+q-k+n)!}{(L+q-k)!n!}~. \qe

\comment{
We summarize the $SU(n+1)$ content of $(0,k)$-forms in the table below:
\begin{center}\begin{tabular}{|c||c|c|}
  \hline \vspace{-.4cm}& & \\
  $(0,k)$-form & $\bar{\D}$-exact & $\bar{\D}$-nonexact \\ \hline \hline
   $(0,0)$ & & $(l+q,0,\cdots,0,1)$ \\ \hline
   $(0,1)$ & $(l+q,0,\cdots,0,1) $ & $(l+q-2,1,0,\cdots,0,l) $\\\hline
$(0,2)$ & $(l+q-2,1,0,\cdots,0,l)$ & $(l+q-3,0,1,\cdots,0,l)$\\\hline $\vdots$ & $\vdots$ & $\vdots$\\ \hline $(0,N)$ & $(l+q-N,0,\cdots,0,l+1)$ &\\ \hline
\end{tabular}
\end{center}
where a sum is implied over l. $\bar{\D}$ takes us from nonexact $(0,k)$-forms to exact $(0,k+1)$-forms and vice versa for $\bar{\D}^*$. In particular $\bar{\D}^*$ vanishes on nonexact forms.
}

We will now write eigenspinors of the Dirac operator. We have seen that the space of $\bar{\D}$-nonexact $(0,k)$-forms, $k=0, \ldots, n-1$ breaks up into irreducible
representations
\eq \sum_{l=0;\;l+q\geqslant k+1}^L \quad\raisebox{9.4mm}{$\overbrace{\overline{\yng(4)}}^l$} \hspace{-0.808mm}\overbrace{\young(\ \ \ \ ,\ione,\itwo,\cdot,\ik)}^{l+q-k}
\qe
of $su(n+1)$. The $l^{\mathrm{th}}$ mode is spanned by the elements
\begin{align}
\bm{\psi}^k_{\gvec_{l+q}}{}^{\vv{\dl}_{l}}=\calP_{\gvec_{l+q}\bvec_l}{}^{\vv{\dl}_l\avec_{l+q}}\Ad_{\al_{l+q}}\cdots\Ad_{\al_{k+2}}\Ad_{[\al_{k+1}}\ket{\vv{\mu}_{L-l}}&\bra{\vv{\mu}_{L-l}
\bvec_l}a^{\i_1}_{\al_1}\cdots a^{\i_k}_{\al_k]}\non \\& \otimes \g^{\bi_1}\cdots \g^{\bi_k}\ket{\Omega}~\non\\
=\calP_{\gvec_{l+q}\bvec_l}{}^{\vv{\dl}_l\avec_{l+q}}\Ad_{[\al_{k+1}}\ket{\al_{k+2} \cdots \al_{l+q}\vv{\mu}_{L-l}}&\bra{\vv{\mu}_{L-l}
\bvec_l}a^{\i_1}_{\al_1}\cdots a^{\i_k}_{\al_k]}\non \\& \otimes \g^{\bi_1}\cdots \g^{\bi_k}\ket{\Omega}~,\label{nonexact(0,k)form}
\end{align}
where the projector $\calP$ removes all $\avec\leftrightarrow \bvec$ contractions and
anti-symmetris\-ation is over the indices $\al_1,\ldots, \al_{k+1}$. In appendix B we  show that
$\bm{\psi}^k_{\avec_{l+q}}{}^{\bvec_{l}}$  is an eigenoperator of $\Dirac^2=\bar{\D}\bar{\D}^\dagger+\bar{\D}^\dagger\bar{\D}=\g^\bi\g^\j \K_\bi \K_j+\g^\j \g^\bi \K_j \K_\bi$
\eq \Dirac^2 \bm{\psi}^k_{\avec_{l+q}}{}^{\bvec_{l}}=\lambda^2 \bm{\psi}^k_{\avec_{l+q}}{}^{\bvec_{l}} \qe
 with non-zero eigenvalues
\eq \lambda^2=(l+q)(l-k+n), \quad l=0\ldots L~~\mathrm{subject~to}~~ l+q\geqslant k+1,  \qe
with $k=0,\ldots,n-1$.
Eigenspinors $\bm{\Psi}^\pm $ are therefore given by
\eq \bm{\Psi}_{\avec_{l+q}}^{\pm,k} {}^{\bvec_{l}}:=\bm{\psi}^k_{\avec_{l+q}}{}^{\bvec_{l}}\pm \frac{1}{\lambda}\Dirac\,\bm{\psi}^k_{\avec_{l+q}}{}^{\bvec_{l}} \qe
with eigenvalues $\pm\lambda$ and degeneracies given by the dimension
(\ref{spinordimensions}) of the
relevant $su(n+1)$ irreducible representations.

These values agree, up to the cut-off at $L$, with the available continuum spectrum and degeneracies given for spin structure on $\CP^n$ \cite{Cahen89:spect_dirac,Cahen94:errat,Bar96:harmon_spinor,SeiSemm:cpn} and spin$^\mathrm{c}$ structures on $\CP^2$ \cite{Pope:1980ub,Grosse:1999ci}. The number of, and conditions for, zeros modes of the Dirac operator, given in (\ref{zeromodesk=1}) and (\ref{zeromodesk=N}) also agrees with the continuum result \cite{Dolan:2002ck}.

Note that the space of spinors for a given $L$ and $q$ is a linear combination of left modules corresponding to different algebras, $\calA_{L+q-k},\,k=0,\ldots,n$.

\section{Eigenvalues of the Dirac operator}

In this appendix we shall calculate the eigenvalues of the square of the Dirac operator. Consider an $l$-th mode nonexact $(0,k)$-form basis element $\bm{\psi}^k_{\gvec_{l+q}}{}^{\vv{\dl}_{l}}$ described in (\ref{nonexact(0,k)form}),
with $l+q+k\ge 1$ and $0\le k \le n-1$.
Observing that $\hat K_i(\bm{\psi}^k_{\gvec_{l+q}}{}^{\vv{\dl}_{l}})=0$ we
find that acting with $\Dirac^2$ gives
\begin{align}
\Dirac^2 \bm{\psi}^k_{\gvec_{l+q}}{}^{\vv{\dl}_{l}}&=\bar{\D}^\dagger \bar{\D}\bm{\psi}^k_{\gvec_{l+q}}{}^{\vv{\dl}_{l}}\nonumber\\
&= \K_\i \K_{\bj} \g^\i \g^{\bj} \bm{\psi}^k_{\gvec_{l+q}}{}^{\vv{\dl}_{l}}  \\
&= \K_\i \K_{\bi} \bm{\psi}^k_{\gvec_{l+q}}{}^{\vv{\dl}_{l}}- \K_{\i} \K_{\bj} \g^\bj \g^\i \bm{\psi}^k_{\gvec_{l+q}}{}^{\vv{\dl}_{l}}~.\nonumber
\end{align}

Taking each term individually:
\begin{align}
\K_\i \K_{\bi}&\bm{\psi}^k_{\gvec_{l+q}}{}^{\vv{\dl}_{l}}=\calP_{\gvec_{l+q}\bvec_l}{}^{\vv{\dl}_l\avec_{l+q}}
A^\dagger_{\eta_1} A^{\eta_2} A^\dagger_{[\al_{k+1}}\ket{\al_{k+2}\cdots \al_{l+q} \vv{\mu}_{L-l}}\non \\ &\qquad\qquad\qquad\qquad\bra{\vv{\mu}_{L-l}\bvec}a^{\i_1}_{\al_1}\cdots a^{\i_k}_{\al_k]}a^{\i}_{\eta_2}(a^\dagger)^{\eta_1}_\i \otimes \g^{\bi_1}\cdots \g^{\bi_{k}} \ket{\Omega} \non\\
&=\K_\i \K_{\bi}\left(\calP_{\gvec_{l+q}\bvec_l}{}^{\vv{\dl}_l\avec_{l+q}} A^\dagger_{[\al_{k+1}}\ket{\al_{k+2}\cdots \al_{l+q} \vv{\mu}_{L-l}}\bra{\vv{\mu}_{L-l}\bvec}\right) a^{\i_1}_{\al_1}\cdots a^{\i_k}_{\al_k]}\non \\&\qquad\qquad\qquad\qquad\qquad \otimes \g^{\bi_1}\cdots \g^{\bi_{k}} \ket{\Omega}\non\\
&{}+\sum_{j=1}^k \calP_{\gvec_{l+q}\bvec_l}{}^{\vv{\dl}_l\avec_{l+q}}A^\dagger_{\al_j} A^{\eta_2} A^\dagger_{[\al_{k+1}}\ket{\al_{k+2}\cdots \al_{l+q} \vv{\mu}_{L-l}}\non\\ &\qquad\quad\qquad \bra{\vv{\mu}_{L-l}\bvec}a^{\i_1}_{\al_1}\cdots a^{\i_{j-1}}_{\al_{j-1}} a^{\i_{j}}_{\eta_2}a^{\i_{j+1}}_{\al_{j+1}}\cdots a^{\i_k}_{\al_k]} \otimes \g^{\bi_1}\cdots \g^{\bi_{k}} \ket{\Omega} \non\\
&= \left( (l+q-k)(l+n)-k\right) \bm{\psi}^k_{\gvec_{l+q}}{}^{\vv{\dl}_{l}}
\end{align}
where we have used (\ref{A commutator}), (\ref{KiKbi on nonsquare})
and antisymmetry of $\al_1 \cdots \al_{k+1}$.

We now see that
\begin{align}
\K_{\i} \K_{\bj}& \g^\bj \g^\i  \bm{\psi}^k_{\gvec_{l+q}}{}^{\vv{\dl}_{l}}=
\sum_{j=1}^k (-1)^{j-1} \K_{\i_j} \K_{\bi} \calP_{\gvec_{l+q}\bvec_l}{}^{\vv{\dl}_l\avec_{l+q}}
A^\dagger_{[\al_{k+1}}\ket{\al_{k+2}\cdots \al_{l+q} \vv{\mu}_{L-l}}\non\\ &\qquad\qquad\qquad\bra{\vv{\mu}_{L-l}\bvec}a^{\i_1}_{\al_1}\cdots a^{\i_k}_{\al_k]} \otimes  \g^{\bi}\g^{\bi_1}\cdots \g^{\bi_{j-1}} \g^{\bi_{j+1}}\cdots \g^{\bi_k} \ket{\Omega} \non\\
&= k  \calP_{\gvec_{l+q}\bvec_l}{}^{\vv{\dl}_l\avec_{l+q}}A^\dagger_{\eta_1} A^{\eta_2} A^\dagger_{[\al_{k+1}}\ket{\al_{k+2}\cdots \al_{l+q} \vv{\mu}_{L-l}}\non \\&\qquad\quad\qquad\quad\qquad\bra{\vv{\mu}_{L-l}\bvec}a^{\i_1}_{\al_1}\cdots a^{\i_k}_{\al_k]}a^{\i}_{\eta_2}(a^\dagger)^{\eta_1}_{\i_1}\otimes \g^{\bi}\g^{\bi_2}\cdots \g^{\bi_{k}} \ket{\Omega} \non\\
&= -k \calP_{\gvec_{l+q}\bvec_l}{}^{\vv{\dl}_l\avec_{l+q}} \left(\dl_{\eta_1} ^{\eta_2}
-A^{\eta_2}A^\dagger_{\eta_1}\right)  A^\dagger_{[\al_{k+1}}\ket{\al_{k+2}\cdots \al_{l+q} \vv{\mu}_{L-l}} \non\\
&  \qquad\quad\bra{\vv{\mu}_{L-l}\bvec}a^{\i_1}_{\al_1}\cdots a^{\i_k}_{\al_k]}
\left( (a^\dagger)^{\eta_1}_{\i_1}a^{\i}_{\eta_2}+\dl^\i_{\i_1}\dl^{\eta_1}_{\eta_2}\right)  \otimes \g^{\bi}\g^{\bi_2}\cdots \g^{\bi_{k}} \ket{\Omega}\non\\
&= -k[(n+1)-(n+1+L+q-k)]\bm{\psi}^k_{\gvec_{l+q}}{}^{\vv{\dl}_{l}}\non\\
&\qquad\quad{}-k \calP_{\gvec_{l+q}\bvec_l}{}^{\vv{\dl}_l\avec_{l+q}} A^\dagger_{[\al_{k+1}}\ket{\al_{k+2}\cdots \al_{l+q} \vv{\mu}_{L-l}} \label{crossterm} \\
&\qquad\qquad\qquad\qquad\bra{\vv{\mu}_{L-l}\bvec}a^{\i_1}_{\al_1}\cdots a^{\i_k}_{\al_k]} \hat {J}_{\i_1}{}^\i\otimes\g^{\bi}\g^{\bi_2}\cdots \g^{\bi_{k}} \ket{\Omega},\non
\end{align}
where we have used the fact that one of the cross-terms vanishes, since
\begin{align}
-A^{\eta_2}A^\dagger_{\eta_1}  & A^\dagger_{[\al_{k+1}}\ket{\al_{k+2}\cdots \al_{l+q} \vv{\mu}_{L-l}} \bra{\vv{\mu}_{L-l}\bvec}a^{\i_1}_{\al_1}\cdots a^{\i_k}_{\al_k]}
(a^\dagger)^{\eta_1}_{\i_1}a^{\i}_{\eta_2} \non\\
=-A^{\eta_2}A^\dagger_{\eta_1} & A^\dagger_{[\al_{k+1}}\ket{\al_{k+2}\cdots \al_{l+q} \vv{\mu}_{L-l}} \bra{\vv{\mu}_{L-l}\bvec}(a^\dagger)^{\eta_1}_{\i_1}a^{\i_1}_{\al_1}\cdots a^{\i_k}_{\al_k]}
a^{\i}_{\eta_2} \non \\
=-A^{\eta_2}A^\dagger_{\eta_1} & A^\dagger_{[\al_{k+1}}\ket{\al_{k+2}\cdots \al_{l+q} \vv{\mu}_{L-l}} \bra{\vv{\mu}_{L-l}\bvec}\hat J^{\eta_1}_{\al_1} a^{\i_2}_{\al_2}\cdots
a^{\i_k}_{\al_k]}  a^{\i}_{\eta_2}
\end{align}
and (\ref{numberrel})
shows that $\calP_{\gvec\bvec}{}^{\vv{\dl}\avec}\Ad_{\eta_1} A^\dagger_{[\al_{k+1}}\ket{\avec \vv{\mu}_{L-l}}\bra{\vv{\mu}_{L-l}\bvec} \hat{J}^{\eta_1}{}_{\al_1]}=0$ since
$\calP_{\gvec\bvec}{}^{\vv{\dl}\avec}$ removes all contractions between upper and
lower indices.
Finally commuting $\hat{J}_{\i_1}{}^\i$ in the last term on the
right hand side of (\ref{crossterm}) through the $a^{\i_j}_{\al_j}$ to use (\ref{Jijsinglet}) and noticing the commutator terms give a factor of $n-(k-1)$ we find
\begin{align}
\K_{\i} \K_{\bj} \g^\bj \g^\i  \bm{\psi}^k_{\gvec_{l+q}}{}^{\vv{\dl}_{l}}
&= -k[(k-L-q)+L+n-(k-1)]\bm{\psi}^k_{\gvec_{l+q}}{}^{\vv{\dl}_{l}}\non\\
&=-k(n-q+1)\bm{\psi}^k_{\gvec_{l+q}}{}^{\vv{\dl}_{l}}
\end{align}

Hence
\begin{align}
\Dirac^2 \bm{\psi}^k_{\gvec_{l+q}}{}^{\vv{\dl}_{l}}=\lambda^2\bm{\psi}^k_{\gvec_{l+q}}{}^{\vv{\dl}_{l}}
\end{align}
where $\lambda^2=(l+q)(l-k+n)$, with $l+q+k\ge 1$ and $0\le k\le n-1$.
As far as we know the spin$^c$ spectrum presented here for $n>2$ is new.

\bibliography{bibfile}
\bibliographystyle{JHEP}



\end{document}